\documentstyle[epsfig]{article}

\begin{document}

\title{ A new method to search for a \\ cosmic ray dipole anisotropy}

\author{S. Mollerach and E. Roulet\\
CONICET, Centro At\'omico Bariloche,\\
Bustillo 9500, 8400, Argentina}

\maketitle

\begin{abstract}
We propose a new method to determine the dipole (and quadrupole) component
of a distribution of cosmic ray arrival directions, which can be applied 
when there is  partial sky coverage and/or inhomogeneous exposure. 
In its simplest version it requires that the exposure only 
depends on the declination, but it can be easily extended
to the case of a small amplitude modulation in right ascension.
The method essentially combines a $\chi^2$ minimization
of the distribution in declination to obtain the multipolar components 
along the North-South  axis and a harmonic Rayleigh analysis for the 
components involving the right ascension direction.
\end{abstract}

\section{Introduction}

The search of a dipolar anisotropy in the arrival directions of  cosmic rays
(CRs) is of crucial importance to better understand the  CR origin, the
distribution of their sources and the way they propagate through the
magnetized galactic medium. Several analyses have looked for a first harmonic
in the right ascension distribution of the arrival directions, and some
positive findings have been reported.   These include some giving few per
thousand
anisotropies at energies $\sim 10^{15}$~eV, others giving percent level
anisotropies at $\sim 10^{16}$~eV (which are however in contradiction with
more recent data from Kascade which obtains upper limits below these
findings) \cite{ho05}, and also a 4\% anisotropy at $10^{18}$~eV has 
been reported by the Agasa collaboration \cite{ha99}.
Due to the small size of the anisotropies observed it is clear that these
measurements are extremely delicate since non-uniformities in 
the sky coverage,
due to e.g. asymmetries of the experimental setup or to 
weather effects not properly accounted for, could mimic
the signal searched.

On the theoretical side, realistic models of CR diffusion in the Galaxy,
including the drift component which actually turns out to give the dominant
contribution to the CR macroscopic currents for energies above that 
of the knee and 
up to the ankle \cite{ca03}, 
predict anisotropies steadily increasing with energy, with a
typical size of $10^{-3}$ at $10^{16}$~eV and reaching the percent level at
$10^{18}$~eV. These predictions of course depend on the details of the regular
and turbulent galactic magnetic field models adopted, on the CR composition
assumed  as well as on the
distribution of cosmic ray sources, which are believed to be galactic below
the ankle.

The Auger experiment will gather the largest CR statistics at EeV energies and
above, and can hence study accurately the large scale anisotropies in these
energy regimes. This should allow to test Agasa's findings at EeV energies and
also to study the ankle region, which is of particular interest because it is 
there
that a transition from a galactic to an extragalactic component is believed to
occur, and hence the behavior of the anisotropies at these energies could
provide important clues to understand this transition.

Besides the 2D harmonic analysis, which determines the phase in right
ascension (in addition to the Rayleigh amplitude), 
it is clearly desirable to reconstruct  
the CR dipole in 3D, i.e. getting its actual amplitude and 
direction in the sky 
(both in right ascension
and declination).  A procedure to achieve this in the case of an experiment
with full sky coverage was put forward by Sommers some time ago
\cite{so01},  and was
recently generalized by Aublin and Parizot \cite{au05}
to the case of partial exposure relevant for present
experiments.
The basis of Sommers' method is to take each event
direction ${\hat n}_i$, 
with an associated exposure $\omega_i$, and obtain the
dipole through
\begin{equation}
\vec D=\frac{3}{\mathcal N}\sum_i\frac{{\hat n}_i}{\omega_i},
\end{equation}
with ${\mathcal N}\equiv \sum_i \omega_i^{-1}$ ($i=1,...N$ labels 
the different events observed).
This approach (and its generalization to partial sky coverage as well, 
hereafter referred to as the SAP method) 
has the virtue of being very easy to implement once the exposure is estimated,
but has the following possible drawbacks:

\begin{itemize}
\item Since each event is weighted by the inverse of its associated exposure,
 different regions
  in the sky do not have their real statistical weight in the determination of
  the dipole, as they have all been artificially uniformised by dividing by
  the exposure. Moreover,
  events in low exposure regions have very large weights and can hence give
  rise to enhanced fluctuations in the results. 
\item The uncertainties in both the amplitude and the direction of the dipole
  are not obtained from the data but rather using Monte Carlo
  simulations with a known dipole anisotropy and comparing the input dipole
parameters with the reconstructed values.
\item The method has no means of diagnosing whether 
 the departure from isotropy has a
  dipolar shape, and for instance an intrinsic quadrupole 
 or even a localized excess in a region of the sky will lead to a
  non-vanishing dipole amplitude. In these situations the uncertainties
  estimated from the MC simulations will clearly have little or no meaning. 
\end{itemize}

\section{The $\chi^2$+Rayleigh method}

We will here consider an alternative method to determine a dipole signature
which tries to overcome some of 
the difficulties just mentioned and apply it to 
simulated data sets.

Let $\hat d$ be the dipole's direction, and $\gamma={\rm acos}({\hat
  n}\cdot \hat d)$ the angle between the dipole and the event's arrival
  direction ${\hat n}$. A dipolar
  distribution should give rise to a CR flux (incident on the Earth) of the
  form
\begin{equation}
\frac{{\rm d}\Phi}{{\rm d}\Omega}\propto (1+a\cos\gamma),
\end{equation}
where $a$ is the dipole's amplitude (i.e. $\vec D=a\hat d$).
When this flux is observed by an experiment with non-uniform exposure
(integrated over time) $\omega(\hat n)$, the expected event rates should
behave as
\begin{equation}
\frac{{\rm d}N}{{\rm d}\Omega}={\mathcal N}\omega({\hat n})(1+a\cos\gamma)
\label{dndom}
\end{equation}
with the normalization, which depends on $a$ and $\hat d$, fixed to
reproduce  the total number of events observed.
Suppose that now we try to fit a dipole signal along a direction 
$\hat d'$, along which the expected distribution would be 
\begin{equation}
\frac{{\rm d}N}{{\rm d}\cos\gamma'}={\mathcal N}\int_0^{2\pi}{\rm d}\theta'\
\omega(\hat n)(1+a\cos\gamma),
\end{equation}
where $\theta'$ is the azimuthal angle around the axis $\hat d'$,
$\cos\gamma'\equiv \hat d'\cdot \hat n$ and
\begin{equation}
\cos\gamma\equiv \hat d\cdot\hat n=
\cos\beta\cos\gamma'+\sin\beta\sin\gamma'\cos(\theta'-\theta_d'),
\label{cosg.eq}
\end{equation}
where $\cos\beta\equiv \hat d\cdot \hat d'$ and $\theta_d'$ is the azimuthal
angle, measured around $\hat d'$, of the dipole vector.

It is then clear that if $\omega$ were uniform in the sky, the term
proportional to $\cos(\theta'-\theta_d')$ in the above integral would 
vanish, leading to
the behavior d$N/{\rm d}\cos\gamma'\propto(1+a\cos\beta\cos\gamma')$,
and hence the dipole's amplitude inferred, barring statistical fluctuations, 
 would be $a\cos\beta$, which is just the dipole component along the $\hat d'$
 axis. One may then envisage that the dipole's direction could be obtained as
 the one maximizing the reconstructed dipole's amplitude, but
however,
for non-uniform exposures the $\cos(\theta'-\theta_d')$ term does not 
average to zero in general, so that this procedure could lead to a biased
result. 

There is however a particularly relevant case where this bias is absent, which
is when one considers $\hat d'=\hat z$, i.e. the NS equatorial axis, since for
the case of uniform exposure in right ascension (with an arbitrary declination
dependence), the $\cos(\theta'-\theta_d')$ term in the
 integral will indeed vanish\footnote{The other situation in which the result
is unbiased is of course when $\hat d'=\hat d$, since in this case
$\beta=0$.},  hence leading to a behavior 
d$N/{\rm d}\cos\gamma'\propto(1+a_z\cos\gamma')$, with
$a_z=a\sin\delta$ being the amplitude of the 
$z$ component of the dipole.  This then suggests to use 
a $\chi^2$ method to fit a dipolar distribution along
the $\hat z$ direction to get an unbiased estimate of $a_z$. 

To fit a dipole signal along a generic direction $\hat d'$, one can just take
 $n_\gamma$ bins of $\cos\gamma'$ (e.g. $n_\gamma=10$ is a reasonable
choice, and we checked that using a larger value does not improve 
significantly the results), and compute the number of events in 
each bin $N_j$ and the
corresponding expected values 
\begin{equation}
{\bar N}_j=\int_{\Delta\Omega_j}{\rm d}\Omega\ \frac{{\rm d}N}{{\rm d}\Omega},
\label{nj}
\end{equation}
where $\Delta\Omega_j$ is the solid angle for which $\cos\gamma'$
 falls in the $j$-th bin, with ${{\rm d}N}/{{\rm d}\Omega}$ given by
 Eq.~(\ref{dndom}). 
We can then write the $\chi^2$ function as
\begin{equation}
\chi^2(a)=\sum_{j=1}^{n_\gamma}{(N_j-{\bar N}_j)^2\over 
{\bar N}_j},
\label{chi2}
\end{equation}
where the sum is clearly restricted to the bins\footnote{In
case one bin ends up with a very small number of events, it may just  be 
convenient to choose a different number of bins for the fit.}
 with ${\bar N}_j>0$. 
One can then determine the value
$a_m$ for which $\chi^2(a)$ is minimized, and obtain
$\Delta a$ such that $\chi^2(a_m+\Delta a)=\chi^2(a_m)+1$. 

We hence apply this method along the $\hat z$ direction to obtain first $a_z$.
Concerning the dipole component orthogonal to the
$\hat z$ axis, the natural strategy is to apply Rayleigh's method.
In particular, it has been noticed in  \cite{au05} that this method gives, 
at least in  the case of partial sky coverage, 
a better determination of the dipole's right ascension 
than the SAP method. 
 One can use then Rayleigh's method  to get 
the best estimate of the dipole's right ascension and first harmonic
amplitude.  
This method\footnote{The Rayleigh method is based on the assumption that the
  experimental exposure is uniform in right ascension. This may require
  eventually to select a subset of the data corresponding to
  sidereal days in which this assumption holds to a specified level of
  accuracy or, as discussed further below, to generalise the method so 
as to include the effects of right ascension non uniformities, if 
these can be properly modelled.}
 consists simply of computing the quantities (here $\alpha$ is the right
 ascension) 
\begin{equation}
A=\frac{2}{N}\sum_{i=1}^N \cos \alpha_i\ \ \ {\rm and}\ \ \ \ B=
\frac{2}{N}\sum_{i=1}^N \sin \alpha_i
\end{equation}
and then get the first harmonic amplitude $r$ and phase $\Psi$
through
\begin{equation}
r=\sqrt{A^2+B^2}\ \ \ {\rm and}\ \ \ \ \Psi={\rm atan}\frac{B}{A}.
\end{equation}
In addition, as shown in ref.~\cite{au05}, there is a simple relation between
$r$ and the original dipole components $a_z$ and 
$a_\perp\equiv a\cos\delta$ in the case in which the exposure is 
independent of $\alpha$,
which is 
\begin{equation}
r=\left|{c_3 a_\perp\over c_1+c_2 a_z}\right|
\label{rvsci.eq}
\end{equation}
where
$$c_1=\int_{\delta_{min}}^{\delta_{max}}{\rm d}\delta\ \omega(\delta)
  \cos\delta\ \ \ \ \ \ $$
\begin{equation}
c_2=\int_{\delta_{min}}^{\delta_{max}}{\rm d}\delta\ \omega(\delta)
  \cos\delta \sin\delta
\label{cis.eq} 
\end{equation}
$$c_3=\int_{\delta_{min}}^{\delta_{max}}{\rm d}\delta\ \omega(\delta)
  \cos^2\delta.\ \ \ \  $$
One can then reconstruct completely the dipole's direction and
  amplitude from the previous expressions. 
Moreover, once the dipole's direction is obtained, it is useful
  to redetermine the dipole's amplitude $a$ using the $\chi^2$
  minimization method for that fixed direction, and in this way one also
evaluates the
  $\chi^2/dof$ for this fit, what provides a check of the dipolar shape 
of the data distribution. 

\section{Results}

As an example, we show the results of applying this method to data sets of
$3\times 10^4$ simulated events with an intrinsic dipole of 5\% amplitude
($a=0.05$) pointing towards $(\delta,\alpha)=(-45^\circ,0^\circ)$, assuming an
experiment at the Auger location (latitude $-35.2^\circ$) with an ideal
geometric exposure (uniform in right ascension and with a zenith angle
dependence d$N/{\rm d}\theta\propto \sin\theta\cos\theta$, assuming a maximum
zenith angle for the events analyzed of 60 degrees).

Let us for instance describe in more detail the results of one particular
simulated data set.
For this simulation the reconstructed dipole 
 has an amplitude $a= 0.059\pm 0.013$ pointing
towards $(\delta,\alpha)=(-52^\circ,-6^\circ)$, which is $\sim 10^\circ$  
degrees apart from the input direction of
the simulations. 
Let us notice that the error $\Delta a$ depends essentially only on the
dipole's declination and the statistics at hand, scaling with 
the overall statistics as
$N^{-1/2}$, as expected. We find indeed that, for the assumed detector's 
location coincident with the Auger latitude and for maximum zenith 
angles of $60^\circ$, the empirical expression 
\begin{equation}
\Delta a\simeq \sqrt{3\over N}\left( 1+0.6\sin^3 |\delta|\right)
\end{equation}
reproduces quite reasonably the results of many simulations
performed with different input parameters.

\begin{figure}[ht]
\centerline{{\epsfig{width=3.in,file=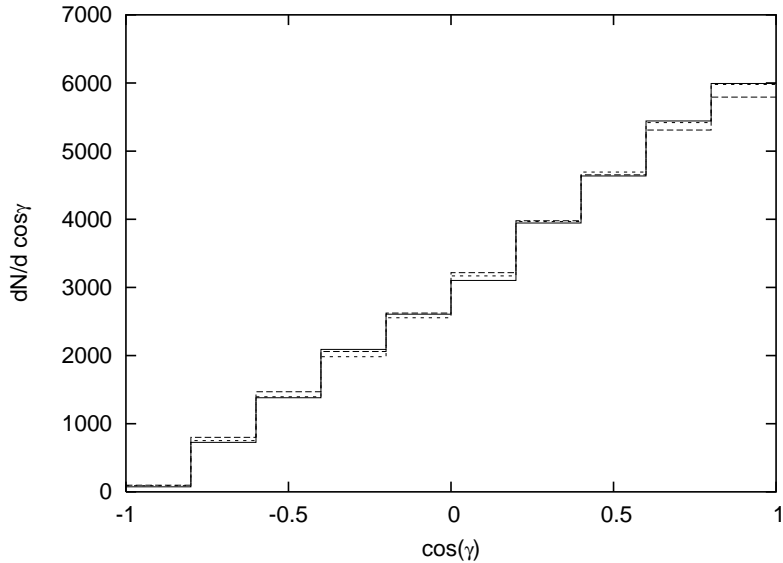,angle=-90}}}
\caption{Histogram of events vs. $\cos\gamma$ (solid), of isotropic
  expectations (dashed) and best dipole fit (dotted) for the simulation
  described in the text. }
\label{fitdip}
\end{figure}

Fig.~\ref{fitdip} shows the distribution 
of the events (solid line), the expectations of an isotropic sky (dashed line) 
and the best fit obtained through the method outlined (dotted line),
as a function of 
$\cos\gamma\equiv\hat d\cdot \hat n$.

\begin{figure}[ht]
\centerline{{\epsfig{width=3.in,file=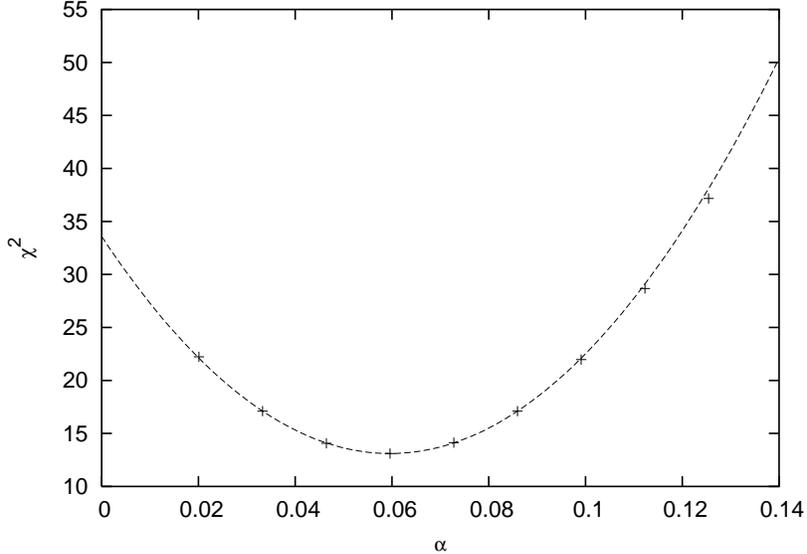,angle=-90}}}
\caption{$\chi^2$ values for the direction maximizing the dipole's amplitude
  and fitting parabolic function. }
\label{chi2.fig}
\end{figure}

The $\chi^2$ value of
the best fit is here 13.1, 
and notice that the number of degrees of freedom is just
$n_\gamma'-1$, where $n_\gamma'$ is the number of bins which are not empty 
(which can be less than $n_\gamma$ due to the partial sky coverage, and 
moreover in this last case it will depend on the dipole's orientation).
For this example $n'_\gamma=10$, giving a value for $\chi^2/dof=1.46$, 
which is not unreasonable. A plot of the $\chi^2$ 
function for the best fit direction, as well as the fitting function 
$\chi^2(a)=\chi^2(a_m)+((a-a_m)/
\Delta a)^2$, are displayed in Fig.~\ref{chi2.fig}.

Making this analysis for a set of 500 simulations like the one just described, 
we got the amplitude, right ascension and declination
values (average and dispersion) which are displayed in Table~1.
It is apparent from this Table that both methods give average values quite
close to the input ones, so that their biases are small, and the
$\chi^2+Rayleigh$ method leads to dispersions\footnote{Clearly part of the
  discrepancy between the input and reconstructed values is just due to
  statistical fluctuations in the simulated data, while another piece will be
  associated to the reconstruction method itself.} in the recovered values 
smaller by about 20\%  than SAP's method. This is probably related to 
the possible large enhancement of the effects of
statistical fluctuations in the low exposure regions near the boundaries 
of the observed sky, which affect  the
Aublin-Parizot generalization of Sommers' method for
partial sky coverage.

\begin{table}[ht]
\centerline{\begin{tabular}{|c||c|c||c|c||c|c|}
\hline
\ & $\langle a\rangle$ [\%]&$\sigma_a$ & $\langle\delta\rangle$& 
$\sigma_\delta$ & $\langle \alpha\rangle$ & 
$\sigma_{\alpha}$  \\
\hline
 \  & \ & \ & \ & \ & &\\
SAP & 5.8 & 1.7 & $-44.6^\circ$&$ 19.1^\circ$  & $-0.3^\circ$&$
21.6^\circ$  \\
 \  & \ & \ & & & &  \\
\hline
 \  & \ & \ & & & & \\
$\chi^2+ {\rm Rayleigh}$  & 5.4& 1.3 & $-44.5^\circ$&$ 16.1^\circ$  & 
$0.2^\circ$&$ 17.6^\circ$   \\
 \  & \ & \ & & & & \\
\hline
\end{tabular}}
\caption{Results from MC simulations (as described in the text) for the
  average values and dispersion in the reconstructed 
dipole's amplitude, declination and right ascension. The input values used
were $a_{input}=5$\%, $\delta=-45^\circ$ and $\alpha=0^\circ$.}
\end{table}

In  the last row of Table~1 we display 
the values of the dipole amplitude obtained as $\sqrt{a_z^2+a_\perp^2}$, but
for  comparison the values of 
$a$ which results from the $\chi^2$ minimization once the direction of the 
dipole has been fixed using $a_z$, $r$ and $\Psi$, lead to an average
value $\langle a_m\rangle=5.4\pm 1.4$\%, and the $\chi^2$ values of the 
corresponding fits are characterized by 
$\langle\chi^2/dof\rangle=0.88\pm 0.45$.

Another advantage of the method proposed here is that it really exploits the
dipolar shape of the signal searched. For instance, we performed 200 MC 
simulations
with $3\times 10^4$ events each 
with a distribution consisting of an isotropic background
plus a quadrupole, taken for definiteness with symmetry of revolution around
an axis $\hat q$,
i.e. such that
\begin{equation}
\frac{{\rm d}\Phi}{{\rm d}\Omega}\propto \left(1+\frac{Q}{2}\left[3(\hat n\cdot
\hat q)^2-1\right]\right). 
\label{quadrupole}
\end{equation}

We adopted a quadrupole  amplitude $Q=0.1$ because this value leads to an
inferred dipole $a\simeq 0.05$, as in the examples discussed previously,
and the orientation $\hat q$ adopted also points towards
$(\delta,\alpha)=(-45^\circ,0^\circ)$. 
Applying  the method introduced above 
to reconstruct a dipole leads to
 $\langle a\rangle = 6.3\pm 1.1$\% (with  $\langle a_m\rangle = 
5.3\pm 1.1$\%), but the key point is that the 
best fit results have associated values of  $\langle\chi^2/dof\rangle=3.5\pm 
1.3$, 
which are quite poor, and hence this configuration cannot be mistaken 
with a dipolar one. 
Applying the
SAP method to these simulations 
gives reconstructed dipole's amplitudes with 
 $\langle a\rangle=4.2\pm 1.5$\% 
and the
associated uncertainty in $a$ that would have been estimated for this case 
is just the same as given in Table~1, i.e. $\Delta a\simeq 1.7$\%.

\section{Incorporating right ascension modulation effects}

A basic underlying assumption of the Rayleigh method is the 
uniformity of the exposure of the experiment with respect 
to right ascension.
This however does not always hold, due e.g. to power or to 
communication failures, so that the strategy usually followed
in order to apply this method is to first select a subset of the
data corresponding to whole sidereal days in which the running
conditions of the experiment under consideration were stable.
This procedure can however significantly reduce the amount of 
data available. For instance, the Kascade Collaboration \cite{an04} was 
left with only $\sim 20$~\% of their original data when this 
kind of selection was performed.

We will here introduce a procedure which allows to take into 
account right ascension modulation effects, due to e.g. non-uniform
running conditions of the experiment, making possible hence to use
 the whole statistics to obtain the amplitude and phase
of the right ascension modulation of the incident cosmic ray flux.

Suppose one has obtained, by properly taking into account the detector's
down time \cite{ha05}, the right ascension dependent experimental exposure 
$\omega(\delta,\alpha)$.
This allows then to introduce a declination only dependent average 
exposure through
\begin{equation}
\bar\omega(\delta)=\frac{1}{2\pi}\int_0^{2\pi}{\rm d}\alpha\ \omega(\delta,\alpha).
\end{equation}
It is now possible to generalize Rayleigh's method by introducing
\begin{equation}
\tilde{A}\equiv \frac{2}{\tilde N}\sum_i \cos\alpha_i
\frac{\bar\omega(\delta_i)}{\omega(\delta_i,\alpha_i)},
\end{equation}
and
\begin{equation}
\tilde{B}\equiv \frac{2}{\tilde N}\sum_i \sin\alpha_i
\frac{\bar\omega(\delta_i)}{\omega(\delta_i,\alpha_i)},
\end{equation}
where
\begin{equation}
\tilde{N}\equiv \sum_i 
\frac{\bar\omega(\delta_i)}{\omega(\delta_i,\alpha_i)}.
\end{equation}

Notice that now the contribution of each event to $\tilde{A}$, $\tilde B$ 
and to $\tilde N$ is weighted by the factor 
$\bar\omega(\delta_i)/\omega(\delta_i,\alpha_i)$, which
 takes into account the right ascension non-uniformities, but being
 of order unity does not overweight the contribution of events in regions
of  relatively low exposure.

By analogy with the standard approach, a modified Rayleigh amplitude 
and phase  can now be introduced through
\begin{equation}
\tilde\Psi={\rm atan}\frac{\tilde{B}}{\tilde{A}}\ \ \ ,\ 
\ \ \tilde r=\sqrt{\tilde{A}^2+\tilde{B}^2}.
\end{equation}

It is possible to generalize now eqs.~(\ref{rvsci.eq}) 
and (\ref{cis.eq}) by identifying
\begin{eqnarray}
\tilde{A}& =& \frac{2}{\tilde N}\int {\rm d}\Omega\ \omega(\delta,\alpha)
\frac{{\rm d}\Phi}{{\rm d}\Omega}(\delta,\alpha)\cos\alpha 
\frac{\bar\omega(\delta)}{\omega(\delta,\alpha)}\cr
&=&
\frac{2}{\tilde N}\Phi_0\int {\rm d}\delta\ \bar\omega(\delta)\cos\delta
\int{\rm d}\alpha\ \cos\alpha\left[ 1+a_\parallel\sin\delta+
a_\perp\cos\delta\cos(\alpha-\alpha_d)\right] \cr
&=& \frac{2\pi}{\tilde N}\Phi_0 \bar{c}_3a_\perp\cos\alpha_d
\end{eqnarray}
where we assumed that the cosmic ray flux is dipolar, i.e. following 
$\Phi=\Phi_0(1+a\hat d \cdot\hat n)$, and $a_\parallel=a_z$ while 
$a_\perp$ is the amplitude of the dipole vector in the $xy$ plane, with
$\alpha_d$ being the right ascension of the dipole vector.

Similarly one can find that
\begin{equation}
\tilde B= \frac{2\pi}{\tilde N}\Phi_0 \bar{c}_3a_\perp\sin\alpha_d
\end{equation}
and
\begin{equation}
\tilde N=2\pi\Phi_0(\bar c_1+a_\parallel \bar c_2)
\end{equation}
with
\begin{eqnarray}
\bar c_1&=&\int {\rm d}\delta\  \bar\omega(\delta)\ \cos\delta \cr
\bar c_2&=&\int {\rm d}\delta\  \bar\omega(\delta)\ \cos\delta\sin\delta\cr
\bar c_3&=&\int {\rm d}\delta\  \bar\omega(\delta)\ \cos^2\delta.
\end{eqnarray}
One then finds that
\begin{equation}
\tilde r=\left|{\bar c_3a_\perp\over \bar c_1+\bar c_2 a_\parallel}\right|
\end{equation}
while the Rayleigh phase $\tilde \Psi$ turns out to be just the right ascension
of the dipole $\alpha_d$.

\section{Including the quadrupole}

One of the main concerns in the reconstruction of a given multipole
from data of a partially covered sky, as for example the dipole
reconstruction discussed at the beginning, 
is the possible mixing with the other
multipoles. For example, as we have shown, a non-vanishing
quadrupole can lead to a dipolar signal when only a region of the sky
is observed with an inhomogeneous exposure. A first control on the
true dipolar character of the large scale anisotropy is given by  the
value of the 
$\chi^2/dof$. A more careful analysis can be performed by
reconstructing the next multipole, namely the quadrupole, and
quantifying in this way 
their relative strength and the amount of leaking among them.

We propose here an extension of the $\chi^2$+Rayleigh method to reconstruct a
dipolar $+$ quadrupolar signal. This distribution would give rise to a 
CR flux
\begin{eqnarray}
\frac{{\rm d}\Phi}{{\rm d}\Omega} =\Phi_0\left( 1+a_z \sin\delta +a_x
\cos\delta \cos\alpha +a_y \cos\delta \sin\alpha + 
\frac{Q_{zz}}{2}
\sin^2\delta\right. \cr
 +\frac{Q_{xx}}{2}\cos^2\delta \cos^2\alpha
+ \frac{Q_{yy}}{2}\cos^2\delta \sin^2\alpha +
Q_{xy}\cos^2\delta
\sin\alpha \cos\alpha \cr
\left. + Q_{xz}\cos\delta \sin\delta \cos\alpha
+ Q_{yz}\cos\delta \sin\delta \sin\alpha \right),
\label{d+q}
\end{eqnarray}
where we have taken the $\hat z$ axis along the north pole direction,
and the quadrupole tensor is symmetric and traceless.
The observed flux, assuming for simplicity that the exposure only
depends\footnote{The extension for $\omega(\delta,\alpha)$
  is straightforward along the lines proposed in the previous section.}
 on $\delta$, is given by
\begin{equation}
\frac{{\rm d}N}{{\rm d}\Omega}={\mathcal N}\omega(\delta)
\frac{{\rm d}\Phi}{{\rm d}\Omega},
\end{equation}
where the normalization, which now depends on the amplitude and
orientation of the dipole and quadrupole, is fixed to reproduce the
total number of events.

The first step is again to fit the distribution of events in $\delta$
integrated over $\alpha$, taking advantage of the fact that the exposure is
uniform in $\alpha$. Now this distribution will be a
function of not only $a_z$, but also of $Q_{zz}$, namely
\begin{equation}
\frac{{\rm d}N}{{\rm d}\sin\delta} \propto
\left(1-\frac{Q_{zz}}{4}+a_z \sin\delta+ \frac{3Q_{zz}}{4}\sin^2\delta\right).
\end{equation}
The values of $a_z$ and $Q_{zz}$ are obtained from a $\chi^2$
minimization similar to that performed in Eqs.~(\ref{nj},\ref{chi2})
(notice that $\cos\gamma'=\sin\delta$ for an axis along the $\hat z$ 
direction). 
 
The next step is to use a generalized Rayleigh's method including
higher order harmonics in order to obtain the rest of the dipole and
quadrupole components. This is performed by computing the quantities
\begin{eqnarray}
 A &=&{2\over N}\sum_{i=1}^N \cos \alpha_i\ ,\ \ \ \ \ \ \ \ \ \  B =
{2\over N}\sum_{i=1}^N \sin \alpha_i\ , \cr
C&=&{2\over N}\sum_{i=1}^N \cos \alpha_i \sin \alpha_i\ , \ \   D =
{2\over N}\sum_{i=1}^N \sin^2 \alpha_i\ , \cr
E&=& {2\over N}\sum_{i=1}^N \cos \alpha_i \cos \delta_i\ , \ \   F =
{2\over N}\sum_{i=1}^N \sin \alpha_i  \cos \delta_i\ .
\end{eqnarray}

These quantities can be easily related to the multipole coefficients
in Eq.~(\ref{d+q}) using $\sum_i f(\delta_i,\alpha_i) = \int {\rm d}\Omega
\  \omega(\delta){\rm d}\Phi/{\rm d}\Omega\;f(\delta,\alpha)$. We thus obtain
\begin{eqnarray}
A&=&\frac{2 \pi \Phi_0}{N} \left(a_x\; \overline{\cos\delta} + Q_{xz}\;
\overline{\cos \delta \sin \delta}\right)\cr
B&=&\frac{2 \pi \Phi_0}{N} \left(a_y\; \overline{\cos\delta} + Q_{yz}\;
\overline{\cos \delta \sin \delta}\right)\cr
C&=&\frac{\pi \Phi_0}{2 N} Q_{xy}\;
\overline{\cos^2 \delta}\cr
D&=&\frac{2 \pi \Phi_0}{N} \left(\overline{1} + a_z\; \overline{\sin\delta} 
+ Q_{zz} \left(\frac{1}{2}\;\overline{\sin^2 \delta}
-\frac{1}{8}\;\overline{\cos^2 \delta} \right) +\frac{1}{4}Q_{yy}\;
\overline{\cos^2 \delta}\right)\cr
E&=&\frac{2 \pi \Phi_0}{N} \left(a_x \;\overline{\cos^2\delta} + Q_{xz}\;
\overline{\cos^2 \delta \sin \delta}\right)\cr
F&=&\frac{2 \pi \Phi_0}{N} \left(a_y \;\overline{\cos^2\delta} + Q_{yz}\;
\overline{\cos^2 \delta \sin \delta}\right)
\end{eqnarray}
Finally, the total number of events is related to $\Phi_0$, $a_z$ and
$Q_{zz}$ through 
$$N=2 \pi \Phi_0 \left(\overline{1} + a_z\; \overline{
\sin\delta} + \frac{Q_{zz}}{4}\left(3\; \overline{\sin^2\delta}-
\overline{1}\right) \right).$$
In the previous expression we have  used the notation
$$\overline{f(\delta)} = \int_{\delta_{min}}^{\delta_{max}} {\rm d} \delta
\cos \delta \omega(\delta) f(\delta).$$

The remaining dipolar and quadrupolar coefficients are then obtained
as
\begin{eqnarray}
a_x &=& \frac{{\mathcal{K}}_1}{{\mathcal{K}}_2}
\left(A\; \overline{\cos^2 \delta \sin \delta} 
- E\; \overline{\cos \delta \sin \delta}\right)\cr
a_y &=& \frac{{\mathcal{K}}_1}{{\mathcal{K}}_2}
\left(B\; \overline{\cos^2 \delta \sin \delta} 
- F\; \overline{\cos \delta \sin \delta}\right)\cr
Q_{xz} &=& \frac{{\mathcal{K}}_1}{{\mathcal{K}}_2}
\left( E\; \overline{\cos \delta}
-A\; \overline{\cos^2 \delta}\right) \cr
Q_{yz} &=& \frac{{\mathcal{K}}_1}{{\mathcal{K}}_2}
\left(  F\; \overline{\cos \delta}
-B\; \overline{\cos^2 \delta} \right)\cr
Q_{xy} &=& \frac{4 }{\overline{\cos^2 \delta}}\;C {\mathcal{K}}_1\cr
Q_{yy} &=& \frac{4}{\overline{\cos^2 \delta}}\left({\mathcal{K}}_1 D -
\overline{1}-a_z\; \overline{\sin \delta} - \frac{Q_{zz}}{8}\left(4 \;
\overline{\sin^2 \delta}-\overline{\cos^2
  \delta}\right)\right)\cr
Q_{xx} &=& -Q_{yy} - Q_{zz},
\end{eqnarray}
with 
\begin{eqnarray}
{\mathcal{K}}_1&=&\frac{N}{2\pi\Phi_0}=\overline{1}+a_z\; \overline{
  \sin \delta} +\frac{Q_{zz}}{4}(3\; \overline{\sin^2 \delta} -\overline{1}),
\cr 
{\mathcal{K}}_2&=& \overline{\cos \delta}\ \  \overline{\sin\delta 
\cos^2 \delta}
 -\overline{\cos^2 \delta}\ \ \overline{\sin\delta \cos \delta}.
\end{eqnarray}

As an example, we generated two datasets of $4\times 10^5$ events each, 
one with a dipole and the other with a quadrupole. 
For the first, we took a dipole amplitude $a=0.1$  pointing along
$(\delta,\alpha)=(-45^\circ,0^\circ)$. Applying the method introduced 
above, we found from the $\chi^2$ minimization\footnote{In this case, 
since we have to determine two parameters, it is convenient to
take a larger number of bins of $\cos\gamma$, e.g. $n_\gamma=20$.}
 along the $z$-axis that $a_z=-0.085\pm 0.011$ and $Q_{zz}=-0.018\pm 0.017$.
The generalized Rayleigh method then gave: $a_x=0.075$, $a_y=-0.005$, 
$Q_{xx}=0.007$, $Q_{yy}=0.011$, $Q_{xy}=-0.004$, $Q_{xz}=0.002$ and
$Q_{yz}=-0.004$, in perfect agreement with the input values, which
correspond to $(a_x,a_y,a_z)=(0.0707,0,-0.0707)$ and $Q_{ij}=0$.
Notice that applying to this same dataset the $\chi^2+Rayleigh$ method
introduced initially, i.e. without allowing for a quadrupole contribution, 
leads to $a_z=-0.075\pm 0.005,\ a_\perp=0.072$, with $a=0.104\pm 0.004$, 
and pointing only $2^\circ$ away from the input direction.

Regarding the dataset with an input quadrupole, for which we took the
expression given in Eq.~(\ref{quadrupole}), with $Q=0.1$ and $\hat q$ also 
pointing towards 
$(\delta,\alpha)=(-45^\circ,0^\circ)$, we got from the $\chi^2$ minimization
that $a_z=-0.014\pm 0.010$ and $Q_{zz}=0.045\pm 0.016$. For the other 
components we obtained 
 $a_x=0.010$, $a_y=-0.002$, 
$Q_{xx}=0.055$, $Q_{yy}=-0.100$, $Q_{xy}=0.000$, $Q_{xz}=-0.129$ and
$Q_{yz}=-0.011$, in perfect agreement with the input values, which
for the quadrupole assumed correspond to $Q_{zz}=Q/2$, $Q_{xx}=Q/2$,
$Q_{yy}=-Q$ and $Q_{xz}=-3Q/2$, with all other parameters vanishing.

\section{Summary}

In this work we have introduced a method to obtain large scale anisotropies 
in cosmic ray arrival directions in three dimensions. In its simplest 
version, the method allows to obtain the dipole vector describing
the anisotropy of the incident flux, assuming that
the exposure is uniform in right ascension, as approximately holds for 
surface arrays.  The $z$ component $a_z$ is obtained from a $\chi^2$ fit
to the event distribution along the NS axis, and the perpendicular
component is obtained with a 2D Rayleigh analysis in the orthogonal plane.
We compared  this method with the one recently introduced by Aublin and 
Parizot, showing that the dispersions in the values obtained are typically
20\% smaller, and also the $\chi^2$ fit along the reconstructed dipole's
direction provides a measure of the agreement between the arrival
direction distribution and a dipolar shape.

We then generalized the method to the case in which a known modulation,
e.g. induced by detector's down time, induces non-uniformities of the 
exposure with respect to right ascension. Finally, we extended the method
to also include a quadrupole anisotropy, and showed with simulated datasets
that this allows to recover the three components of the dipole and 
the 6 independent components of the quadrupole reliably.

The methods here introduced  should be applicable to detectors like Auger, 
and should help determine the large scale patterns of the cosmic
ray distribution, which can provide crucial information about the  origin
of the cosmic rays
  and the way they propagate through the Galaxy.

\section*{Acknowledgments}
This work is partially supported by CONICET, 
Fundaci\'on Antorchas and ANPCyT.
We are grateful to D. Harari, A. Letessier Selvon, E. Parizot and 
P. Sommers for useful discussions.


\begin{thebibliography}{99}


\bibitem{ho05} For a recent discussion, see J. Hoerandel, astro-ph/0501251.

\bibitem{ha99} N. Hayashida et al. (AGASA Collaboration) 
ICRC 1999, Salt Lake City, OG.1.3.04, astro-ph/9906056.

\bibitem{ca03} J. Candia, S. Mollerach and E. Roulet, JCAP {\bf 05} (2003) 003.

\bibitem{so01} P. Sommers, Astropart. Phys. {\bf 14} (2001) 271.

\bibitem{au05} J. Aublin and E. Parizot, ``Generalized 3D reconstruction
  method of a dipole anisotropy in cosmic-ray distributions'', 
astro-ph/0504575.

\bibitem{an04} T. Antoni et al. (Kascade collaboration) ApJ {\bf 604} 
(2004) 687.

\bibitem{ha05} J. Ch. Hamilton et al., in preparation.

\end{thebibliography}
\end{document}